\documentclass{cs20proc}

\usepackage{kantlipsum}
\usepackage[counterclockwise]{rotating}
\usepackage{multirow}

\editors{S. J. Wolk}
\publisher{Zenodo}
\conference{The 20th Cambridge Workshop on Cool Stars, Stellar Systems, and the Sun}
\conferencedate{2018}

\title{Exploring the Influence of Density Contrast on Solar Near-Surface Shear}
\author{Loren Matilsky,$^{1,2}$ 
        Bradley W. Hindman$^{1,2}$
    	and Juri Toomre$^{1,2}$}

\affiliation{$^{1}$ JILA \\
			 $^{2}$ Department of Astrophysical \& Planetary Sciences, University of Colorado, Boulder, CO 80309-0440}

\shorttitle{Influence of Density Contrast on Near-Surface Shear}
\shortauthors{Matilsky et al.}

\abs{The advent of helioseismology has determined in detail the average rotation rate of the Sun as a function of radius and latitude. These data immediately reveal two striking boundary layers of shear in the solar convection zone (CZ): a tachocline at the base, where the differential rotation of the CZ transitions to solid-body rotation in the radiative zone, and a 35-Mm-thick near-surface shear layer (NSSL) at the top, where the rotation rate slows by about 5\% with increasing radius. Though asteroseismology cannot probe the differential rotation of distant stars to the same level of detail that helioseismology can achieve for the Sun, it is possible that many cool stars with outer convective envelopes possess similar differential rotation characteristics, including both a tachocline and a NSSL. Here we present the results of 3D global hydrodynamic simulations of spherical-shell convection for a Sun-like star at different levels of density contrast across the shell. The simulations with high stratification possess characteristics of near-surface shear, especially at low latitudes. We discuss in detail the dynamical balance of torques giving rise to the NSSL in our models and interpret what these balances imply for the real Sun. We further discuss the dynamical causes that may serve to wipe out near-surface shear at high latitudes, and conclude by offering some theories as to how this problem might be tackled in future work.}

\begin{document}

\maketitle

\section{Introduction}
Helioseismology has revealed the average rotation rate of the Sun's interior to great accuracy. These data indicate two striking boundary layers of shear in the solar convection zone (CZ). In the tachocline at the base of the CZ, there is a sharp transition between differential rotation (fast equator, slow poles) to solid-body rotation in the radiative zone. At the top of the CZ, there is a 35-Mm-thick near-surface shear layer (NSSL), in which the average rotation rate of the plasma slows by about 5\% with increasing radius.

\citet{Foukal75} postulate that the NSSL arises from fast, small-scale, rotationally-unconstrained fluid parcels conserving angular momentum in their radial motion just underneath the photosphere. This process, they argue, tends to homogenize angular momentum in the Sun's near-surface layers and thus creates a negative radial gradient in the rotation rate. Previous numerical simulations of rotating, global, spherical-shell convection have confirmed that a near-surface layer of rotationally-unconstrained flows can be generated in models with high enough density contrast across the shell (e.g., \citealt{Gastine13}; \citealt{Guerrero13}; \citealt{Hotta15}). In particular, \citet{Hotta15} find that strong shear in their model forms at low latitudes, with weak shear at high latitudes.

Here we simulate global, spherical-shell, hydrodynamic convection with a range of density contrasts across the layer. We present results from two models---one with contrast exp(3) $\approx$ 20 and one with contrast exp(5) $\approx$ 150. We henceforth refer to these models as N3 and N5, respectively. For both cases N3 and N5, we have determined the average torque balance in detail and explain how these torques are created. We find, similar to \citet{Hotta15}, that small-scale, fast flows near the outer surface deplete angular momentum from the near-surface layers. However, this is insufficient to create substantial shear, especially at high latitudes.  We discuss this result in the context of the global meridional circulation, which is the mechanism that tends to wipe out shear at high latitudes in our simulations.

\begin{figure*}
	\includegraphics[width=1.05\textwidth]{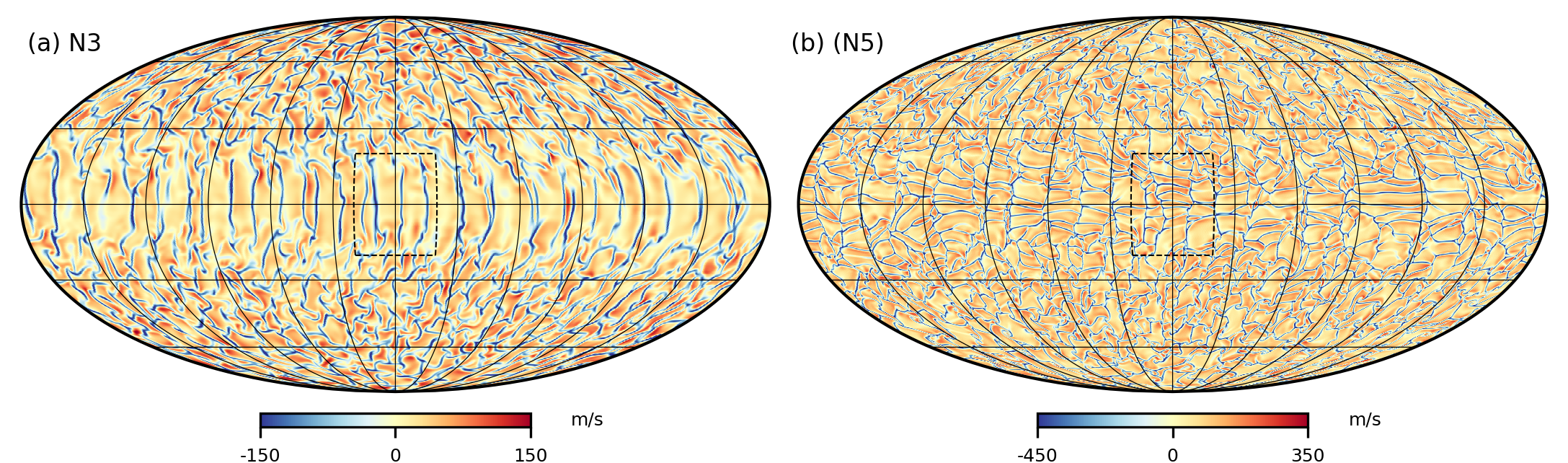}
	\caption{(\textit{a}) and (\textit{b}): Mollweide projections of vertical velocity $v_r$ on full spheres near the outer surface ($r/r_o=0.988$) for cases N3 and N5, respectively. Upflows ($v_r>0$) are shown in red and downflows ($v_r<0$) are shown in blue. In panel (\textit{b}), the colorbar is binormalized to show the asymmetry in the upflow- and downflow-speeds in case N5.}
	\label{fig:sslice}
\end{figure*}
\begin{figure}
	\includegraphics{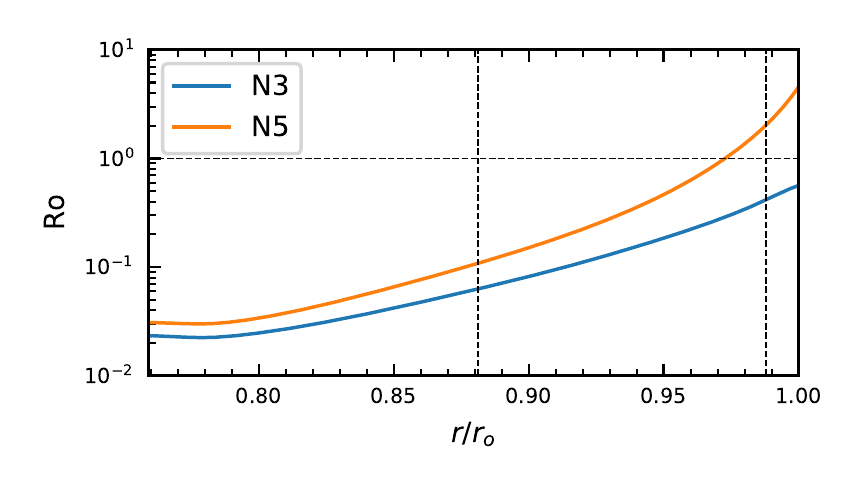}
	\caption{Radial profile of the Rossby number defined in Equation \ref{eq:Roc} for cases N3 and N5. The vertical lines show the depth midway through the shell and the near-surface depth of the global flows shown in Figure \ref{fig:sslice}. The horizontal line shows the critical value of the Rossby number $\mbox{Ro}=1$. \label{fig:rossby_profile} }
\end{figure}

\section{Numerical Experiment}
We use the 3D, open-source, MHD code Rayleigh 0.9.1 \citep{Featherstone16, Matsui16, Featherstone18}, published under the GPL3 license. Rayleigh solves the fully nonlinear equations of viscous hydrodynamics in rotating spherical shells under the anelastic spherical harmonic approach (e.g., \citealt{Clune99}). All models rotate at roughly three times the solar Carrington rate ($ \Omega_0 = 7.8\times10^{-6}\ \mbox{rad}\ \mbox{s}^{-1}$) and have a solar luminosity driven through the shell via a fixed, spherically symmetric internal heating profile. We use a spherically symmetric, temporally steady, adiabatic reference state for the background thermodynamic stratification (the same as described in \citealt{Jones11}). Our spherical shell has inner radius $r_i = 5\times10^{10}$ cm and outer radius $r_o = 6.586\times10^{10}$ cm, corresponding to the lower three density scale heights of the solar convection zone (we do not change the shell depth between cases N3 and N5). 

Explicitly, Rayleigh solves the following equations representing mass, momentum and energy conservation:
\begin{equation}
\nabla\cdot(\overline{\rho}\mathbf{v}) =  0,
\label{eq:cont}
\end{equation}
\begin{eqnarray}
\overline{\rho}\Bigg{[}\frac{\partial\mathbf{v}}{\partial t} + (\mathbf{v}\cdot\nabla)\mathbf{v}\Bigg{]} = -2\overline{\rho}\mathbf{\Omega}_0\times\mathbf{v} \nonumber\\
-\overline{\rho}\nabla \Bigg{(}\frac{P}{\overline{\rho}}\Bigg{)} -\frac{\overline{\rho} S}{c_p}\mathbf{g}
+ \nabla\cdot \mathbf{D}
\label{eq:mom}
\end{eqnarray}
and
\begin{eqnarray}
\overline{\rho}\overline{T}\Bigg{[}\frac{\partial S}{\partial t} + \mathbf{v}\cdot\nabla S\Bigg{]} = \nabla\cdot\big{[}\kappa\overline{\rho}\overline{T}\nabla S\big{]} \nonumber \\
+ 2\overline{\rho}\nu\Big{[}e_{ij}e_{ij} - \frac{1}{3}(\nabla\cdot\mathbf{v})^2\Big{]} + Q,
\label{eq:en}
\end{eqnarray}
respectively. Here $\mathbf{v} = v_r \hat{\mathbf{e}}_r + v_\theta \hat{\mathbf{e}}_\theta + v_\phi\hat{\mathbf{e}}_\phi$ is the fluid velocity in the rotating frame,  $\mathbf{\Omega}_0 = \Omega_0\hat{\mathbf{e}}_z$ is the vector angular velocity, $c_p = 3.5\times 10^8$ erg $\rm{K}^{-1}\ \rm{g}^{-1}$ is the specific heat at constant pressure and $\mathbf{g}$ is the local gravitational acceleration due to a solar mass $M_\odot$ at the origin. The kinematic viscosity $\nu$ and thermometric conductivity $\kappa$ are chosen to be constants in space, equal to $2\times10^{12}$ $\rm{cm}^2\ \rm{s}^{-1}$. These values are several orders of magnitude greater than the molecular diffusivities in the Sun and represent the effects of sub-grid turbulent motions that computationally cannot be resolved. We denote the pressure, density, temperature and entropy by $P$, $\rho$, $T$ and $S$, respectively. We use overbars on the thermodynamic variables to denote the fixed reference state and the lack of overbars to denote deviations from the reference state. The standard Newtonian viscous stress tensor and rate-of-strain tensor  are denoted by $\mathbf{D}$ and $e_{ij}$, respectively. For repeated indices, the standard regular convention is used. 

\section{Flow Structure}
In Figure \ref{fig:sslice} we show the radial velocity just below the outer surface for case N3 (low density contrast) and case N5 (high density contrast). For case N3, flow structures are oriented parallel to the rotation axis at low latitudes, while for case N5, the flows are rapid and isotropic, relatively uninfluenced by the rotation. The columnar sites of upflows and downflows in case N3 correspond to adjacent rolls of fluid aligned with the rotation axis called \textit{Busse columns} (\citealt{Busse02}). Because of their curvature with latitude, they are also known as \textit{banana cells} in the literature.

To quantify the lack of rotational constraint in the highly stratified case N5, we define a \textit{Rossby number} based on the convective flow speeds to be the ratio of rotation period to convective overturning time. The radial profile of the Rossby number can be written
\begin{equation}\label{eq:Roc}
\mbox{Ro}(r) \equiv \frac{v^\prime(r)}{2\Omega_0 H_\rho(r)}.
\end{equation}
Here we have taken the typical convective length scale to be the local density scale height, $H_\rho(r) \equiv - 1/(d\ln{\overline{\rho}}/dr)$. We define the \textit{convective velocity} to be $\mathbf{v}^\prime \equiv \mathbf{v} - \langle\mathbf{v}\rangle$ (the full velocity with the zonally averaged component subtracted out) and the typical convective speed $v^\prime(r)$ to be the rms amplitude of the convective velocity averaged over a spherical surface of radius $r$. 

We plot the radial profile of the Rossby number for cases N3 and N5 in Figure \ref{fig:rossby_profile}. In the deep layers, where $\mbox{Ro}<1$, the level of rotational constraint is high and roughly equal for both cases N3 and N5. Near the outer surface, on the other hand, case N3 is rotationally constrained ($\mbox{Ro}<1$), but case N5 is rotationally unconstrained. In Figure \ref{fig:sslice}, this manifests as the fast, small-scale fluid motions (especially those of the downflows) present in case N5 near the outer surface. 

\section{Differential Rotation}
\begin{figure}
	\includegraphics{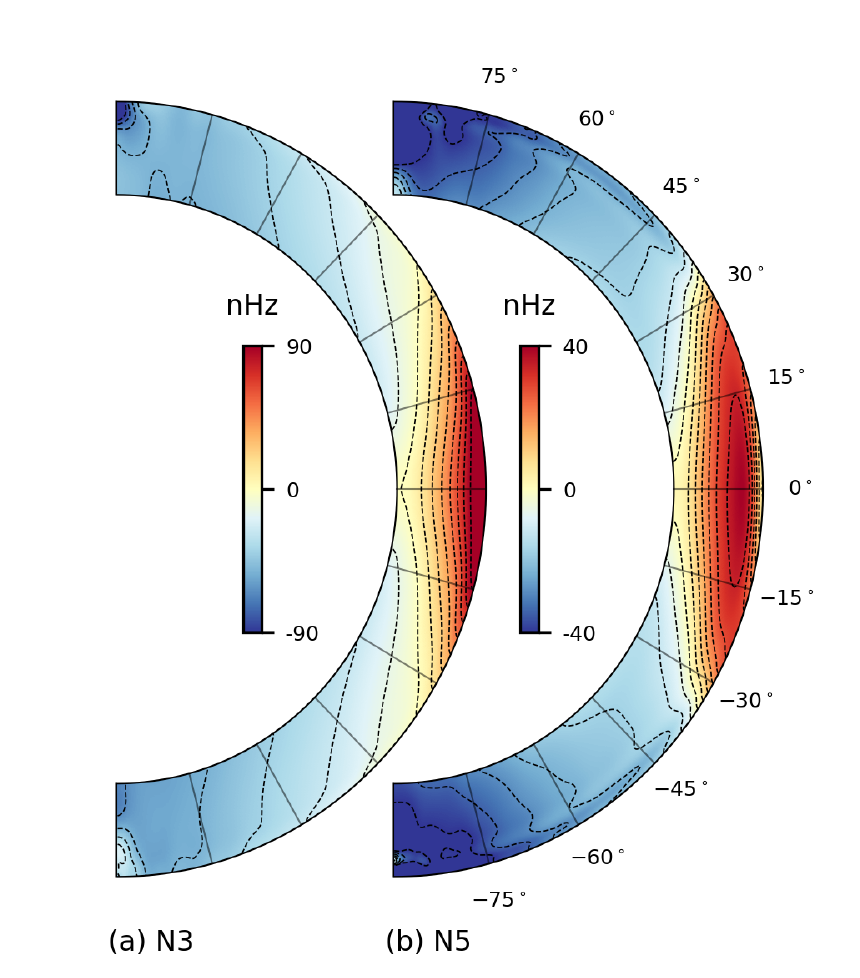}
	\caption{(\textit{a}) and (\textit{b}) show the temporally and zonally averaged angular velocity for cases N3 and N5, respectively. The angular velocity is computed in the rotating frame, so that red tones (positive values) indicate prograde rotation and blue tones (negative values) indicate retrograde rotation. In each panel, 15 contours evenly divide the range of angular velocity indicated by the colorbar. \label{fig:diffrot_merplane}}
\end{figure}

In Figure \ref{fig:diffrot_merplane}, we show the average profile of rotation rate in the meridional plane for cases N3 and N5. While the rotation rate in case N3 increases monotonically with distance from the rotation axis, case N5 exhibits some interesting features reminiscent of near-surface shear. The most prominent of these is a \textit{dimple} in the rotation rate. At low latitudes near the outer surface, the rotation rate decreases with radius by about 3\%. The effect is slightly weaker in magnitude compared to the near-surface shear in the Sun, which is characterized by a roughly 5\% decrease in rotation rate near the top of the CZ.

At high latitudes, there are some signs of shear as well, although the overall effect is much weaker than it is near the equator, with a reduction in angular velocity with radius of only 0.5\%. Furthermore, the shear has both a negative and positive radial gradient, with a small rise in rotation rate (as radius increases), followed by a stronger  dip and then a rise again up to the outer boundary. The profile as a whole is similar to that of \citet{Hotta15}. The dimple, on the other hand, is a a robust feature seen in many simulations (e.g., \citealt{Brun02}; \citealt{Brandenburg07}; \citealt{Guerrero13}). We now seek to understand the features of near-surface shear in case N5 as the dynamical consequence of two types of flow structures: Busse columns and \textit{downflow plumes}.
\begin{figure}
	\includegraphics{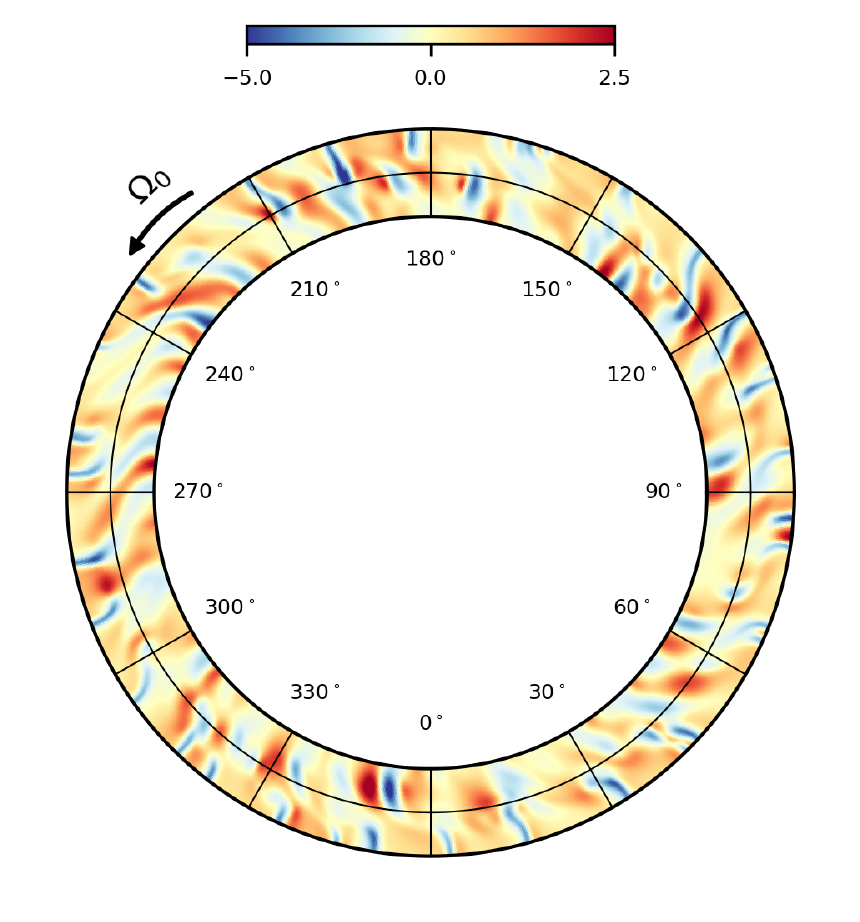}
	\caption{Vertical velocity $v_r$ on the equatorial plane for case N3. Upflows are shown in red and downflows in blue. The velocity has been divided by vertical speed's rms amplitude at each radius, and the colorbar has been binormalized to show the greater speeds of the downflows compared to the upflows. The view is from the North pole, with $\phi$-coordinate increasing in the anticlockwise direction. \label{fig:eqslice} }	
\end{figure}
\section{Busse Columns and Downflow Plumes}
Busse columns are convectively-driven, nonlinear, Rossby waves that appear when the rotational constraint is large (Ro < 1). The Busse-column structure is best seen by examining equatorial cross-sectional cuts of the flow velocity. In Figure \ref{fig:eqslice}, we show the radial velocity on an equatorial slice for case N3 (although Busse columns are also present in case N5, their structure is more clear in case N3). It is clear that the columns have equatorial cross-sections that are tilted such that the top parts are positioned prograde relative to the bottom parts. The orientation of the tilt means that upflows tend to move prograde relative to the background flow and downflows move retrograde, with the net effect that angular momentum is transported outward, increasing the fluid rotation rate far from the rotation axis.

The Busse columns thus create a \textit{solar-like} differential rotation, with an equator (which is farther from the rotation axis) rotating faster than the poles (which are closer to the rotation axis). This effect is clearly illustrated in Figure \ref{fig:diffrot_merplane}a, where the rotation rate in case N3 is roughly constant on cylinders and increases away from the rotation axis. In case N5, this phenomenon is still seen, but is disrupted close to the outer surface where the Rossby number is high. 

\begin{figure}
	\includegraphics{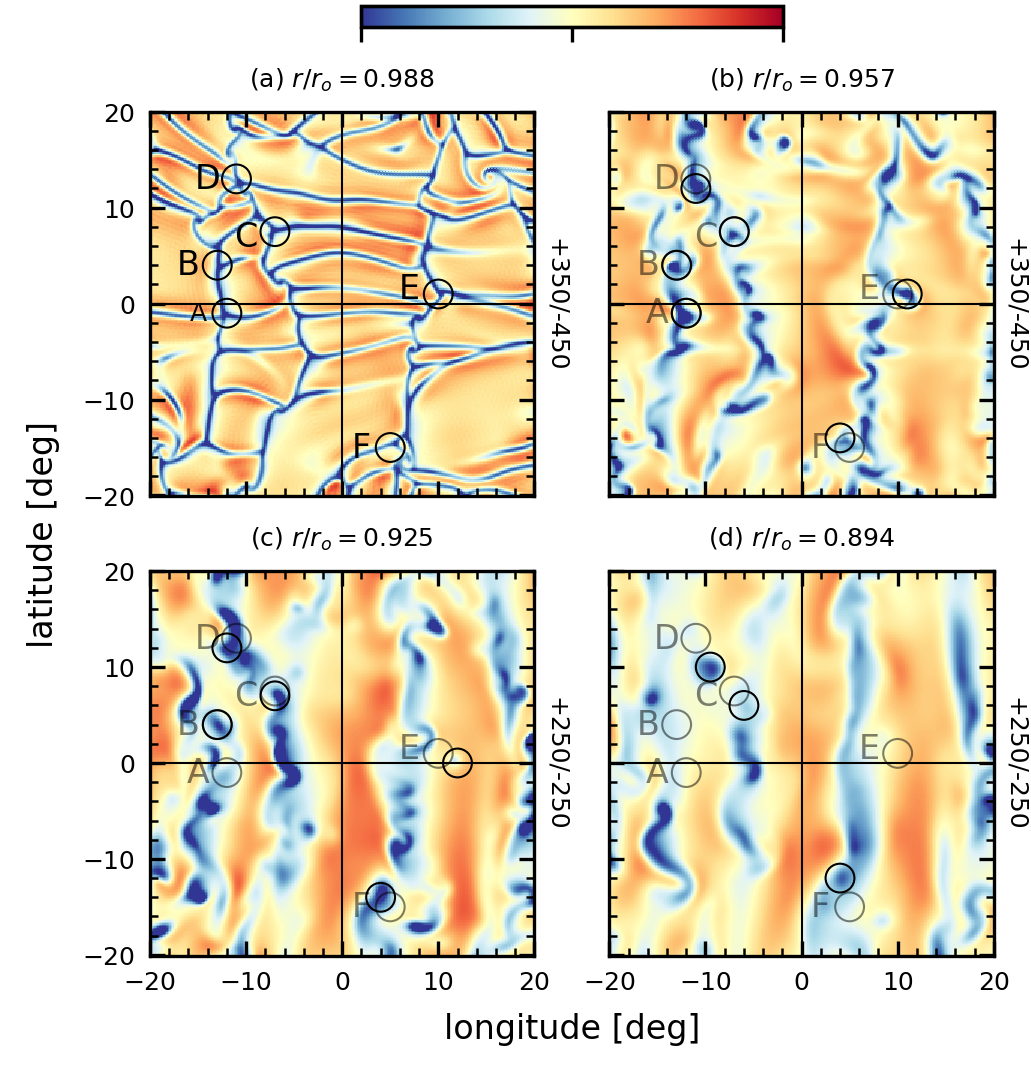}
	\caption{Magnified view of the $40^{\circ}\times40^{\circ}$ rectangular patch of vertical velocity delineated by the dashed box in Figure \ref{fig:sslice}b. Panels (\textit{a})--(\textit{d}) show the same patch at successively deeper layers, evenly spaced between the near-surface layer of Figure \ref{fig:sslice} and mid-depth. The colorbar's saturation values (in m/s) are shown to the right of each panel. The small black circles trace several downflow plumes in depth, labeled in panel (\textit{a}) by capital letters A--F. The near-surface plume locations and labels are shown in gray in later panels. \label{fig:patch_dive}}
\end{figure}

In our models, the high-Rossby-number layer in case N5 comes \textit{primarily from the downflows}. The slow, broad upflows have lower Rossby number and are rotationally constrained. When the fastest downflow structures are traced downward to the deeper layers, we see that they are not entrained in the surrounding columnar structures and instead are in the form of intense regions of downflow with narrow cross-section, which we call \textit{plumes}. The radial structure of the downflow plumes is illustrated in Figure \ref{fig:patch_dive}, where we have magnified a patch of flow close to the equator and followed it radially inward. The upper end of each plume is at a location where a North-West downflow lane crosses an East-West downflow lane. As the depth increases, the plume speeds intensify and the plume itself breaks away from the parent downflow lane in which it is embedded close to the outer surface. The plumes form at all latitudes, and penetrate about halfway through the layer. 

The steady-state differential rotation can be understood in terms of the time-averaged torque balance in the meridional plane. There are three torques that can operate: advection of average angular momentum by the meridional circulation, turbulent flux of angular momentum from velocity correlations (Reynolds stress) and viscous transport of angular momentum. Explicitly, the temporally steady balance of torques in the meridional plane (e.g., \citealt{Elliot00}; \citealt{Brun02}; \citealt{Miesch11}) is given by
\begin{eqnarray}
\tau_{rs} + \tau_{mc} + \tau_v &\equiv& 0, \label{eq:torque_balance}\\
\mbox{where}\ \ \ \ \ \nonumber\\
\tau_{rs} &\equiv& -\nabla\cdot [\overline{\rho} r\sin\theta \langle v^\prime_\phi \mathbf{v}^\prime_m\rangle]\nonumber\\
\tau_{mc} &\equiv& -\langle\overline{\rho}\mathbf{v}_m\rangle \cdot\nabla\mathcal{L}\nonumber\\
\mbox{and}\ \ \ \ \ \tau_{v} &\equiv& \nabla\cdot[\overline{\rho}\nu r^2\sin^2\theta\nabla\Omega]. \label{eq:torque_def}
\end{eqnarray}
Here $\mathbf{v}_m \equiv v_r\hat{\mathbf{e}}_r + v_\theta\hat{\mathbf{e}}_\theta$ is the meridional part of the fluid velocity and $\mathcal{L}\equiv r\sin\theta(\Omega_0r\sin\theta + \langle v_\phi \rangle )=\Omega r^2\sin^2\theta$ is the fluid's specific angular momentum in the non-rotating lab frame. The angular brackets indicate a combined temporal and zonal average and the primes indicate deviations from this average. 
\section{Torque Balance}
\begin{figure}
	\includegraphics{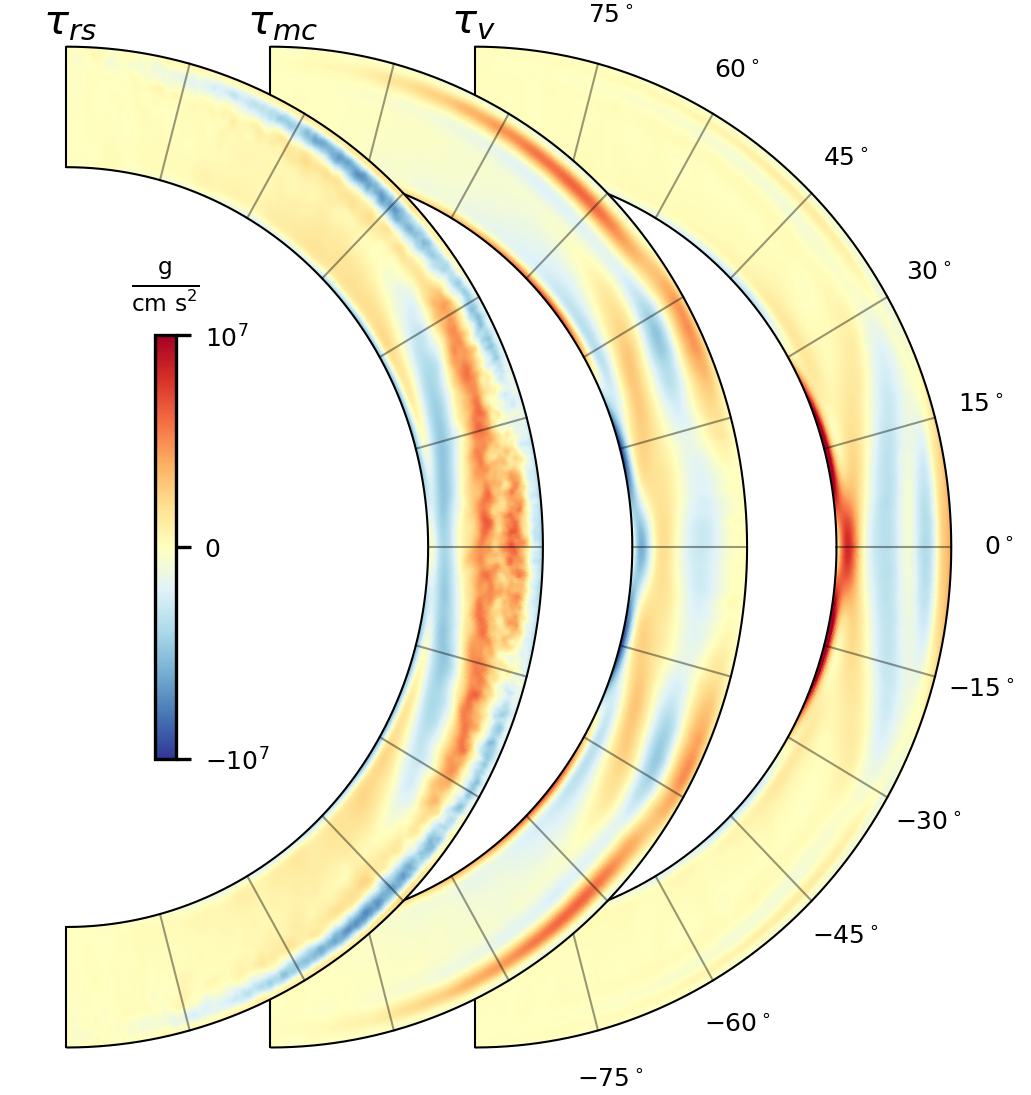}
	\caption{Temporally and zonally averaged torque balance in the meridional plane for case N5. The torque densities due to the Reynolds stress, meridional circulation and viscosity are shown in overlapping plots from left to right. Red tones indicate positive torque and blue tones indicate negative torque. \label{fig:torque_n5}}
\end{figure}
In Figure \ref{fig:torque_n5}, we show the temporally and zonally averaged torque balance in the meridional plane for case N5. At low latitudes, the Reynolds-stress torque is balanced by the viscous torque, whereas at high latitudes, the Reynolds-stress torque is balanced by the meridional-circulation torque. 

The prominent near-surface shear at low latitudes in case N5 can be seen as a direct result of the thin layer of negative Reynolds-stress torque near the outer boundary. This negative Reynolds-stress torque extends to all latitudes and, in fact, has a greater magnitude at high latitudes. This is in juxtaposition to the near-surface shear in case N5, which is very weak at high latitudes. 

At high latitudes, Figure \ref{fig:torque_n5} shows that the balance is primarily between the Reynolds stress and meridional circulation.  From Equation \ref{eq:torque_balance}, this implies $-\tau_{mc}=\tau_{rs}$. Alternatively, using Equation \ref{eq:torque_def}, we find
\begin{equation}
\overline{\rho}\langle\mathbf{v}_m\rangle\cdot\nabla(\Omega r^2\sin^2{\theta})=\tau_{rs}.\label{eq:inertial_balance}
\end{equation}

The Reynolds stress seeks to create near-surface shear at high latitudes in our simulations via the thin band of negative $\tau_{rs}$ near the outer surface. For a relatively featureless meridional circulation cell (i.e., one with little radial variation in the meridional flow), a negative $\tau_{rs}$ would drive near-surface shear by creating a negative radial gradient in $\Omega$ through Equation \ref{eq:inertial_balance}. However, it is apparent from Figure \ref{fig:torque_n5} that the meridional circulation is confined to a thin band of poleward flow near the outer surface---exactly canceling the negative $\tau_{rs}$ with no need for a gradient in $\Omega$. 

In the Sun, there is very little variation in the meridional flow with radius (e.g., \citealt{Giles97}; \citealt{Zhao04}; \citealt{Hathaway12}; \citealt{Chen17}; \citealt{Mandal18}). Thus, the negative Reynolds-stress torque due to the rotationally unconstrained layer hypothesized by \citet{Foukal75} may indeed induce the significant high-latitude shear observed in the Sun. However, it is currently unclear how the meridional circulation in the Sun (or in our case N5, for that matter) is dynamically established. Future work must therefore address in detail the dynamical balance of the meridional circulation achieved in simulations and compare it to the observational deductions from helioseismology. 

\section{Conclusions}
We have performed global, 3D simulations of rotating, spherical-shell convection in order to ascertain the role that the high density contrast across the solar CZ has in maintaining near-surface shear. We have compared one model with low stratification (case N3) and one model with high stratification (case N5) and examined in detail the differential rotation and torque balance established. We have found that:
\begin{enumerate}
	\item Our case N5 possesses a region of rotationally unconstrained fast downflows near the outer surface that transport angular momentum radially inward.
	\item In case N5, the inward transport of angular momentum by the Reynolds stress is balanced by viscous torque at low latitudes and meridional-circulation torque at high latitudes. There is correspondingly substantial shear at low latitudes (though about twofold smaller than that in the Sun) but only very weak shear at high latitudes. 
	\item The absence of substantial high-latitude shear in case N5 is due to the presence of a narrow band of poleward flow in the near-surface layers. This almost completely balances the high-latitude Reynolds-stress torque without the torque forcing a radial gradient in the rotation rate. 
\end{enumerate}
  In conclusion, we have determined that the local transport of angular momentum inward by rotationally unconstrained fluid parcels is insufficient to drive near-surface shear at high latitudes without the presence of a solar-like meridional circulation profile. A more detailed account of this result has been submitted to the Astrophysical Journal \citep{Matilsky18}. The interplay between meridional circulation and differential rotation makes the NSSL a fundamentally \textit{global} problem, requiring numerical simulations in full 3D spherical shells. Future work must address exactly how meridional circulation is dynamically maintained in spherical-shell convection models and ascertain why these profiles are not solar-like. 

\section*{Acknowledgments}
{We thank N. Featherstone for elucidating the interplay between density contrast and rotational constraint, and also for his work as the primary author of the Rayleigh code. We thank K. Julien and M. Miesch for illuminating discussion with regard to the Taylor-Proudman constraint in stratified, spherical-shell convection.
	
This research was primarily supported through NASA grants NNX13AG18G and NNX16AC92G, with additional support through NASA grants NNX14AG05G, NNX14AC05G and NNX17AM01G.
	
The simulations presented in this work were performed on the NASA High-End Computing (HEC) Program through the NASA Advanced Supercomputing (NAS) Division at Ames Research Center.
	
We thank the Computational Infrastructure for Geodynamics (http://geodynamics.org), which is funded by the NSF under awards EAR-0949446 and EAR-1550901, for supporting the Rayleigh code project.
	
L. Matilsky was supported during this work by a Chancellor Fellowship at the University of Colorado Boulder and a George Ellery Hale Graduate Fellowship at the National Solar Observatory.}

\bibliographystyle{cs20proc}
\bibliography{NSSL_CS20_Proceedings.bib}

\end{document}